# Amorphization-Depth-Dependent Bismuth Redistribution and Activation in Silicon for Quantum Devices

Youcef A. Bioud [1], Simon-Jean Michel [2], Sebastien Francoeur [2], Sjoerd Roorda [3], Eva Dupont-Ferrier [1,4]

[1] *Institut Quantique, Université de Sherbrooke, Sherbrooke, Quebec J1K 2R1, Canada*
[2] *Department of Engineering Physics, Polytechnique Montreal, Montreal, QC H3T 1J4, Canada*
[3] *Département de physique, Université de Montréal, Montréal, Québec H3C 3J7, Canada*
[4] *Department of Physics, Université de Sherbrooke, Sherbrooke, Québec J1K 2R1, Canada*

Corresponding authors: y.bioud@usherbrooke.ca , eva.dupont-ferrier@usherbrooke.ca

**Abstract.**

Near-surface bismuth donors are promising for silicon quantum devices because of their rich electron–nuclear spin structure and long spin coherence, but their electrical activation is strongly influenced by implantation-induced amorphization. The redistribution and electrical activation of 20 keV Bi implanted into Si are compared for direct implantation and implantation following Si preamorphization. The two routes produce markedly different amorphous-layer depths of approximately 47 nm for Bi-only implantation and 177 ± 15 nm for Si-preimplanted/Bi. After annealing at 650 °C for 300 s, Rutherford backscattering spectrometry and channeling reveal comparable substitutional Bi fractions of 91 ± 2% for Bi-only and 89 ± 4% for Si-preimplanted/Bi within the main implanted region, whereas Hall measurements indicate Hall-derived electrical activation yields of approximately 60% and 90%, respectively. Depth-resolved ion-scattering analysis reveals features outside the implantation profile, including a buried Bi-related feature at or beyond the initial amorphous/crystalline interface established after implantation, within the end-of-range region, and increasing surface segregation at higher annealing temperatures. The buried feature is consistent with trapping near residual end-of-range defects. X-ray photoelectron spectroscopy and low-temperature photoluminescence measurements on the Bi-only sample reveal no resolvable Bi-oxide contribution and provide a donor-related optical signature, respectively. These results show that the initial amorphization depth and resulting regrowth geometry govern Bi redistribution and influence electrical activation beyond substitutional incorporation alone.

## 1. Introduction.

Group-V donor spins in silicon are promising building blocks for solid-state quantum technologies because they exhibit long coherence times while remaining compatible with established silicon manufacturing and scalable device architectures [1–6]. Atomic-scale fabrication has enabled precise donor positioning[7] , single and coupled-donor transistors [8–11], and electrically controlled donor-based quantum-gate architectures [12–14]. Among group-V impurities, bismuth is particularly attractive because its I=9/2 nuclear spin and strong hyperfine interaction generate a 20-dimensional electron–nuclear Hilbert space [15] that can be exploited for hardware-efficient quantum error-correction schemes [16,17]. Additionally, magnetic-field-insensitive clock transitions, long spin coherence times, electrical spin detection [18], optical addressability, and a GHz-range zero-field splitting favorable for coupling to superconducting resonators further support Si:Bi for quantum memories and hybrid quantum circuits [19–26]. Electrical control and readout generally require donors to be located within a few tens of nanometers of the silicon surface. Ion implantation is well suited to this requirement because it provides controllable implantation depths and compatibility with conventional semiconductor fabrication [27–33]. However, the large atomic mass of Bi produces dense collision cascades and severe lattice disorder. At sufficiently high fluence, the near-surface region becomes amorphous, and thermal annealing restores crystalline order by solid-phase epitaxial regrowth from the underlying silicon substrate [31,34,35]. This process must recover the crystalline lattice, retain Bi near the target depth, incorporate it onto substitutional sites, and suppress electrically inactive or spatially redistributed Bi populations. The electrical activation of implanted Bi may therefore depend not only on its lattice-site fraction but also on the amorphous-layer geometry and the defect environment remaining after regrowth. High substitutional incorporation and electrical activation have been achieved for relatively deep Bi implants. A 150 keV implantation at a fluence of $5 \times 10^{14}\,\mathrm{cm}^{-2}$, followed by annealing at 600 °C for 60 s, produced approximately 95% substitutional incorporation and 90% electrical activation [36]. Shallower implants relevant to quantum devices have generally yielded lower activation efficiencies, particularly when the thermal budget must be restricted to preserve the donor depth distribution and spin properties [27–29,33]. For Bi implanted approximately 20 nm below the surface, Si preamorphization yielded a maximum electrical activation of 64%, despite an almost complete substitutional fraction after low-temperature annealing [31]. This result demonstrated that substitutional occupation alone does not uniquely determine the electrically active donor population. At a Bi fluence of $1 \times 10^{14}\,\mathrm{cm}^{-2}$, direct Bi implantation already amorphizes the near-surface region, whereas an additional Si implant extends the amorphous layer and moves the initial amorphous/crystalline interface farther from the implanted Bi profile [31]. Previous work has mainly correlated activation with annealing conditions, substitutional fraction, and profile broadening [27–29,31,36]. The role of the initial amorphous-layer depth in the spatial partitioning of Bi during and after regrowth remains less clearly established. Bi displaced toward the surface or trapped near the former amorphous/crystalline interface is not fully represented by a substitutional fraction evaluated only within the main implanted region. Residual end-of-range defects may provide preferential sinks for mobile Bi during post-implantation annealing [35]. Resolving these spatially distinct populations is therefore essential for distinguishing lattice-site incorporation from the electrically active fraction of the implanted dose.

Here, we investigate how amorphization depth and solid-phase epitaxial regrowth influence Bi redistribution and electrical activation in near-surface implanted silicon. Bi was implanted at 20 keV with a fluence of $1 \times 10^{14}\ \mathrm{cm}^{-2}$, either directly into silicon or after a Si preimplantation designed to extend the amorphous layer beyond the Bi projected range. The samples were annealed for 300 s between 650 and 950 °C and examined by Hall measurements, secondary ion mass spectrometry, and Rutherford backscattering spectrometry with channeling, complemented by X-ray photoelectron spectroscopy and low-temperature photoluminescence. After annealing at 650 °C, both implantation routes exhibit similarly high substitutional Bi incorporation within the main implanted region, yet the Si-preimplanted/Bi sample shows substantially higher Hall-derived electrical activation, indicating that substitutional occupancy alone does not account for the electrical response. Depth-resolved ion-scattering measurements additionally reveal a buried Bi-related feature located at or beyond the former amorphous/crystalline interface, within the end-of-range region, together with increasing Bi surface accumulation at higher annealing temperatures. The position of this buried feature is consistent with trapping near residual end-of-range defects. These results indicate that the comparable substitutional Bi fractions alone do not explain the different Hall-derived activation yields, and that Bi redistribution into buried and surface-segregated populations may also contribute.

**TABLE 1.** Implantation and annealing conditions for the Bi-only, Si-preimplanted/Bi, and unimplanted Si reference samples.

<table>
<tr><th></th><th>Implant</th><th>Energy (keV)</th><th>Fluence ($\mathrm{cm}^{-2}$)</th><th>Angle (°)</th><th>Annealing temp. (°C)</th><th>Annealing time (s)</th></tr>
<tr><td>Bi-only</td><td>Bi</td><td>20</td><td>$10^{14}$</td><td rowspan="3">7</td><td rowspan="4">650–950</td><td rowspan="4">300</td></tr>
<tr><td rowspan="2">Si-preimplanted/Bi</td><td>Si</td><td>60</td><td>$10^{16}$</td></tr>
<tr><td>Bi</td><td>20</td><td>$10^{14}$</td></tr>
<tr><td>Unimplanted Si reference</td><td>-</td><td>-</td><td>-</td><td>-</td></tr>
</table>

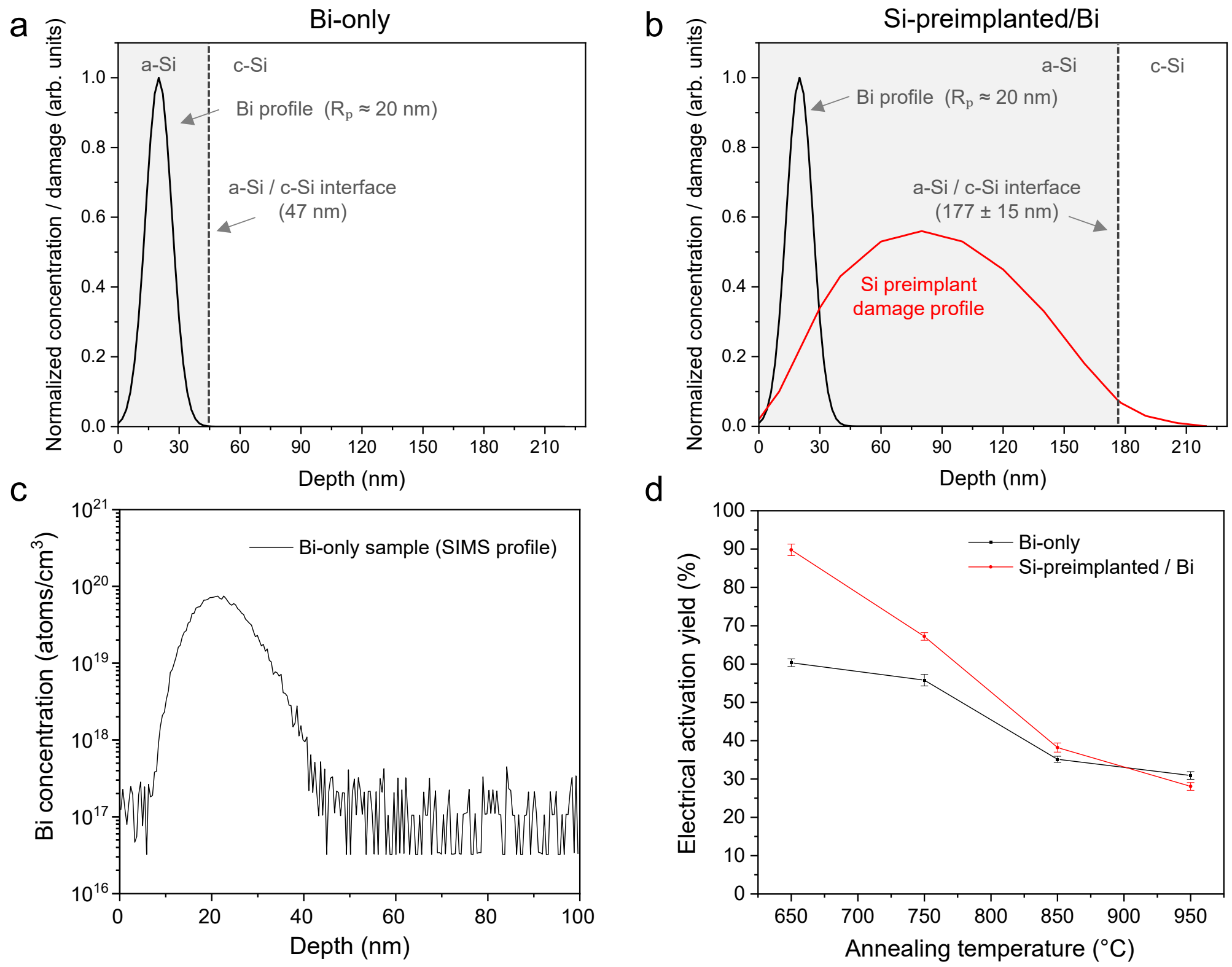

**FIG. 1. Implantation configurations and electrical activation of near-surface Bi in Si. (a,b)** SRIM/TRIM-simulated depth profiles for the Bi-only and Si-preimplanted/Bi routes, respectively. In panel (b), the Bi profile represents the reference SRIM/TRIM profile calculated independently of the preimplantation-induced damage, while the Si preimplantation damage profile is simulated separately. The Bi implantation profile is centered near $R_p \approx 20$ nm, while Si preimplantation produces a substantially deeper damage profile. Dashed lines indicate the corresponding a-Si/c-Si interface positions at approximately 47 nm and 177 ± 15 nm for the Bi-only and Si-preimplanted/Bi samples, respectively. (c) Representative SIMS depth profile of the Bi-only sample before annealing. (d) Hall-derived electrical activation yield as a function of annealing temperature for the two implantation routes. Lines connecting the data points are guides to the eye.

## 2. Results.

High-resistivity, (100)-oriented float-zone Si wafers (4–8 kΩ cm) were divided into two series. The first was implanted with 20 keV Bi ions at a fluence of $1 \times 10^{14}$ cm$^{-2}$(Bi-only), whereas the second received a 60 keV Si$^+$ preimplantation at $1 \times 10^{16}$ cm$^{-2}$ followed by the same Bi implantation (labeled Si-preimplanted/Bi). All implantations were performed at room temperature with a 7° tilt to suppress channeling. The samples were annealed in $N_2$ for 300 s at 650, 750, 850, or 950 °C. Unimplanted wafers subjected to identical anneals served as controls for the Hall measurements. The implantation and annealing conditions are summarized in Table 1. Details of the Hall and Van der Pauw measurement and analysis procedures are provided in ***Supporting Information Section S1***. Bi and implantation-damage profiles were simulated using the Stopping and Range of Ions in Matter (SRIM/TRIM) code [37].

Figures 1(a) and 1(b) compare the simulated implantation profiles for the two routes. The Bi profile shown in both panels corresponds to the reference SRIM/TRIM calculation performed using the same Bi implantation conditions and independently of the preimplantation-induced damage, whereas the Si preimplantation damage profile is simulated separately. The a-Si/c-Si interface positions, approximately 47 nm for Bi-only and 177 ± 15 nm for Si-preimplanted/Bi, were determined experimentally from the as-implanted RBS/channeling spectra. Thus, Si preimplantation introduces a substantially deeper amorphous region and shifts the initial a-Si/c-Si interface farther from the near-surface Bi distribution. Because no SIMS profile is available for the Si-preimplanted/Bi sample, the actual Bi depth distributions of the two routes are not assumed to be strictly identical. This difference in amorphization depth modifies the geometry of solid-phase epitaxial regrowth, consistent with previous channeling studies of near-surface Bi implants [31]. Figure 1(c) shows a near-Gaussian Bi profile centered around 20 nm, with a full width at half maximum of about 14 nm. The retained dose is $9.7 \times 10^{13}\ \mathrm{cm^{-2}}$, within 3% of the nominal fluence. The slightly deeper SIMS peak position relative to the simulated projected range of 17 nm may arise from residual channeling or sputter-depth calibration uncertainty, while no pronounced surface accumulation is observed. Figure 1(d) shows Hall-derived electrical activation yields of approximately 60% for Bi-only and 90% for Si-preimplanted/Bi after annealing at 650 °C for 300 s. These are the highest activation yields observed over the investigated temperature range; the activation subsequently decreases with increasing annealing temperature and converges for the two routes at higher temperatures. The 90% activation obtained for Si-preimplanted/Bi is comparable to that reported for deeper 150 keV Bi implants, where about 95% of the dose was substitutional [36], while our Bi-only and Si-preimplanted/Bi samples exhibit substitutional fractions of 91 ± 2% and 89 ± 4%, respectively, within the main implanted region at 650 °C. Direct comparison, however, should account for differences in implantation depth, preamorphization conditions, and annealing protocol. The activation yield decreases to about 30% for both routes at 950 °C, consistent with the thermal-budget-dependent deactivation reported for deep and near-surface high-fluence Bi implants [31,36]. This loss has been linked to displaced Bi in deeper implants and to reduced substitutional incorporation, diffusion, and inactive Bi-related configurations in near-surface structures [31,36]. Additional electrical transport data are provided in ***Supporting Information Section S1***. The reduced Hall mobility of the implanted samples relative to the unimplanted Si reference is consistent with enhanced impurity and defect scattering. Earlier studies reported increased carrier concentration together with higher resistivity and lower mobility after Bi implantation, with room-temperature mobilities of order $10^{2}\ \mathrm{cm^{2}\,V^{-1}\,s^{-1}}$in heavily doped Si:Bi [30,38]. Although the unimplanted-Si contribution is explicitly subtracted, differences in residual implantation damage may also influence the Hall response through compensation, carrier trapping, mobility variations, or parallel conduction. The values are therefore interpreted as Hall-derived electrical activation yields. The activation trend in Fig. 1(d) cannot be attributed solely to changes in Bi lattice-site occupancy. Instead, it reflects the combined effects of substitutional incorporation, carrier scattering, and Bi redistribution into buried and surface-segregated populations.

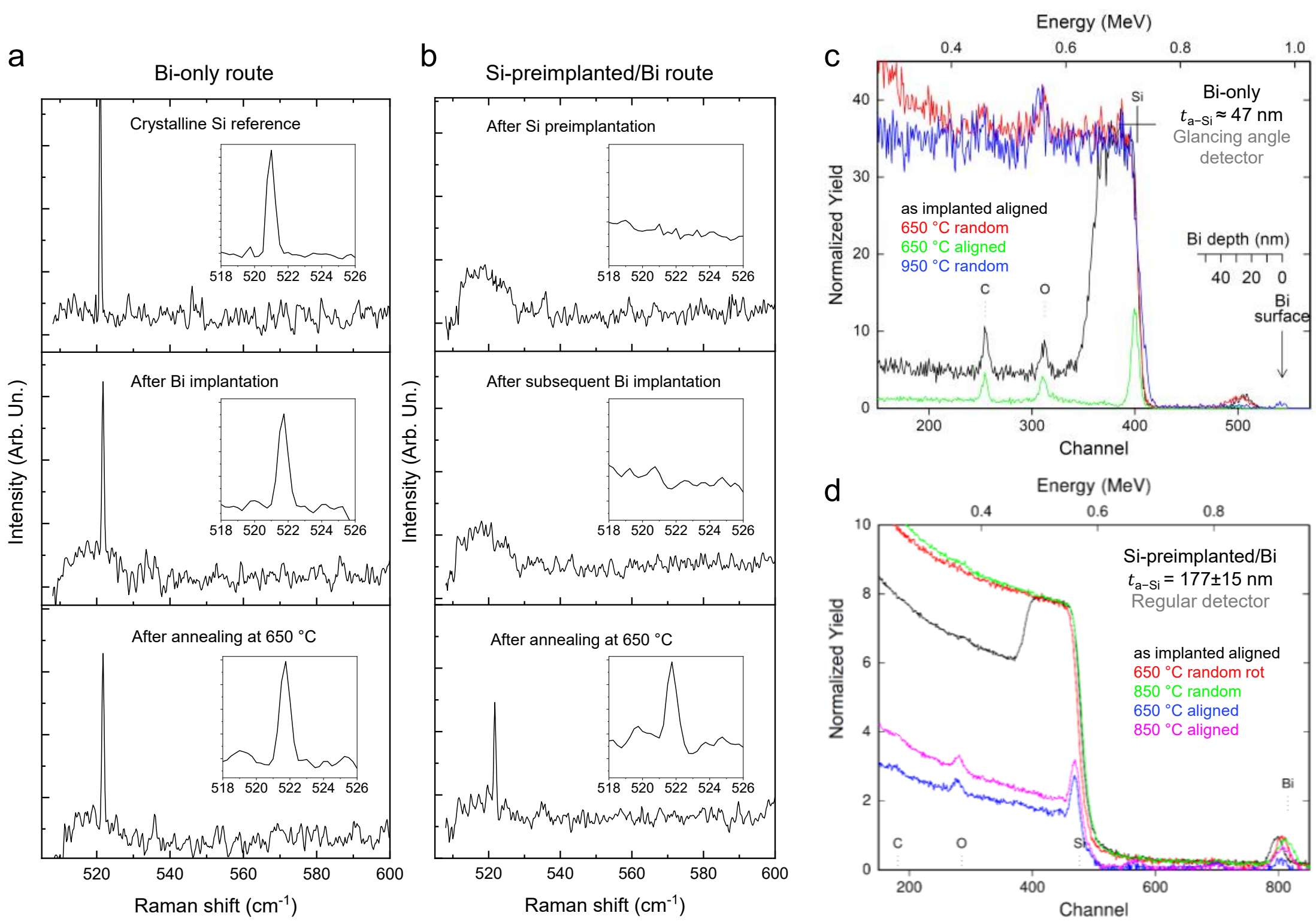

**FIG. 2. Implantation-induced amorphization and lattice recovery in Bi-implanted Si.** (a,b) Raman spectra of the Bi-only and Si-preimplanted/Bi routes at successive processing stages. The reappearance of the crystalline Si peak near 520 cm$^{-1}$ after annealing at 650 °C qualitatively indicates lattice recovery. Insets highlight the spectral region around the crystalline-Si mode near 520 cm$^{-1}$.The Raman spectra are shown over a restricted spectral range to emphasize changes in the crystalline-Si mode. (c,d) Random and aligned RBS spectra of the Bi-only and Si-preimplanted/Bi samples, respectively, including the as-implanted state and selected post-annealing conditions. For each RBS panel, the upper spectrum set corresponds to the glancing-angle detector and the lower set to the regular detector. The glancing-angle geometry provides enhanced depth sensitivity, while the regular detector provides the conventional backscattering geometry. Analysis of the as-implanted aligned spectra gives amorphous-layer thicknesses of approximately 47 nm for Bi-only and 177 ± 15 nm for Si-preimplanted/Bi.

To identify the structural origin of this route-dependent activation, Fig. 2 compares implantation-induced disorder and lattice recovery for the two-sample series. In Fig. 2(a), the top spectrum of the crystalline Si reference shows a sharp mode near 520 cm$^{-1}$, whereas the middle spectrum after Bi implantation exhibits a broader disorder-related spectral contribution. In Fig. 2(b), Si preimplantation produces a predominantly amorphous response, which remains after the subsequent Bi implantation. The stronger suppression of the crystalline-Si contribution in the Si-preimplanted/Bi sample is consistent with its substantially thicker amorphous layer, which more strongly attenuates the excitation and Raman signal from the underlying crystalline substrate. Following annealing at 650 °C, the crystalline-Si mode is largely restored for both routes, indicating substantial lattice recovery, although a weak residual disorder-related contribution remains visible. Because the Raman signal includes a contribution from the underlying crystalline substrate, these spectra are interpreted qualitatively rather than used to determine a crystalline volume fraction.

The aligned RBS spectra provide a depth-sensitive measure of lattice disorder for the two implantation routes. In the Bi-only sample [Fig. 2(c)], the as-implanted spectrum reveals a shallow amorphous layer approximately 47 nm thick. By contrast, Si preimplantation extends the amorphous region to 177 ± 15 nm [Fig. 2(d)], well beyond the near-surface Bi profile. This behavior is consistent with previous channeling measurements showing that a high-fluence Bi implant can itself amorphize near-surface Si, while an additional preamorphization implant displaces the amorphous/crystalline interface deeper into the substrate [31]. The marked reduction of the aligned Si yield after annealing is consistent with extensive lattice recovery by solid-phase epitaxial regrowth in both structures. Additional details of the RBS/channeling analysis are provided in ***Supporting Information Section S2***.

Thus, the two implantation routes recover crystalline order at 650 °C but begin with substantially different amorphous-layer geometries. In the Bi-only sample, the regrowth interface is located at approximately 47 nm, whereas in the Si-preimplanted/Bi sample it is much deeper, at 177 ± 15 nm. The activation contrast in Fig. 1(d) therefore cannot be explained by the presence or absence of recrystallization alone. Instead, the different regrowth distances and defect configurations provide distinct pathways for Bi incorporation and redistribution, including possible interaction with residual end-of-range defects near the former amorphous/crystalline interface [29]. Before examining the depth-resolved Bi populations, we assess the chemical state of Bi and donor-related optical activity in the Bi-only sample after annealing at 650 °C. Figure 3 presents the chemical and optical characterization of the Bi-only sample annealed at 650 °C for 300 s. The depth-resolved Si $2p$ and O $1s$ spectra in Figs. 3(a) and 3(b) show that the oxidized-Si contribution is largely confined to the surface and decreases rapidly after sputtering. Beneath this surface layer, the Si $2p$ response is dominated by elemental Si, with only minor depth-dependent changes that may reflect residual implantation disorder or sputter-induced modification.

The Bi $4f$ core level is sensitive to the local chemical and electronic environment of Bi. Oxidized Bi species are generally shifted toward higher binding energies relative to non-oxidized Bi, whereas the binding energies reported for different non-oxidized Bi configurations can substantially overlap. Consequently, Bi $4f$ binding energies alone cannot uniquely distinguish substitutional from other non-oxidized configurations, since core-level shifts may also reflect differences in local bonding and charge state [39,40]. The Bi $4f$ spectra in Fig. 3(c) show weak features near 156.7 and 162.4 eV, assigned to the Bi $4f_{7/2}$ and Bi $4f_{5/2}$ components, respectively [39]. The signal is most clearly detected between approximately 11 and 28 nm depth, consistent with the SIMS profile shown in Fig. 1(c). No distinct contribution attributable to $Bi_2O_3$ is resolved [41], indicating that oxidation is not the dominant Bi chemical state within the implanted region. However, the low signal intensity and overlap with reported binding energies of elemental Bi prevent XPS from distinguishing substitutional Bi from other non-oxidized configurations [39,40]. The measured Bi $4f$ binding energies are therefore consistent with predominantly non-oxidized Bi within the implanted region, but do not provide an independent determination of its lattice-site occupancy. XPS therefore confirms Bi retention within the intended near-surface region but does not determine its lattice-site occupancy. Additional depth-resolved XPS compositional and core-level data are provided in ***Supporting Information Section S3***.

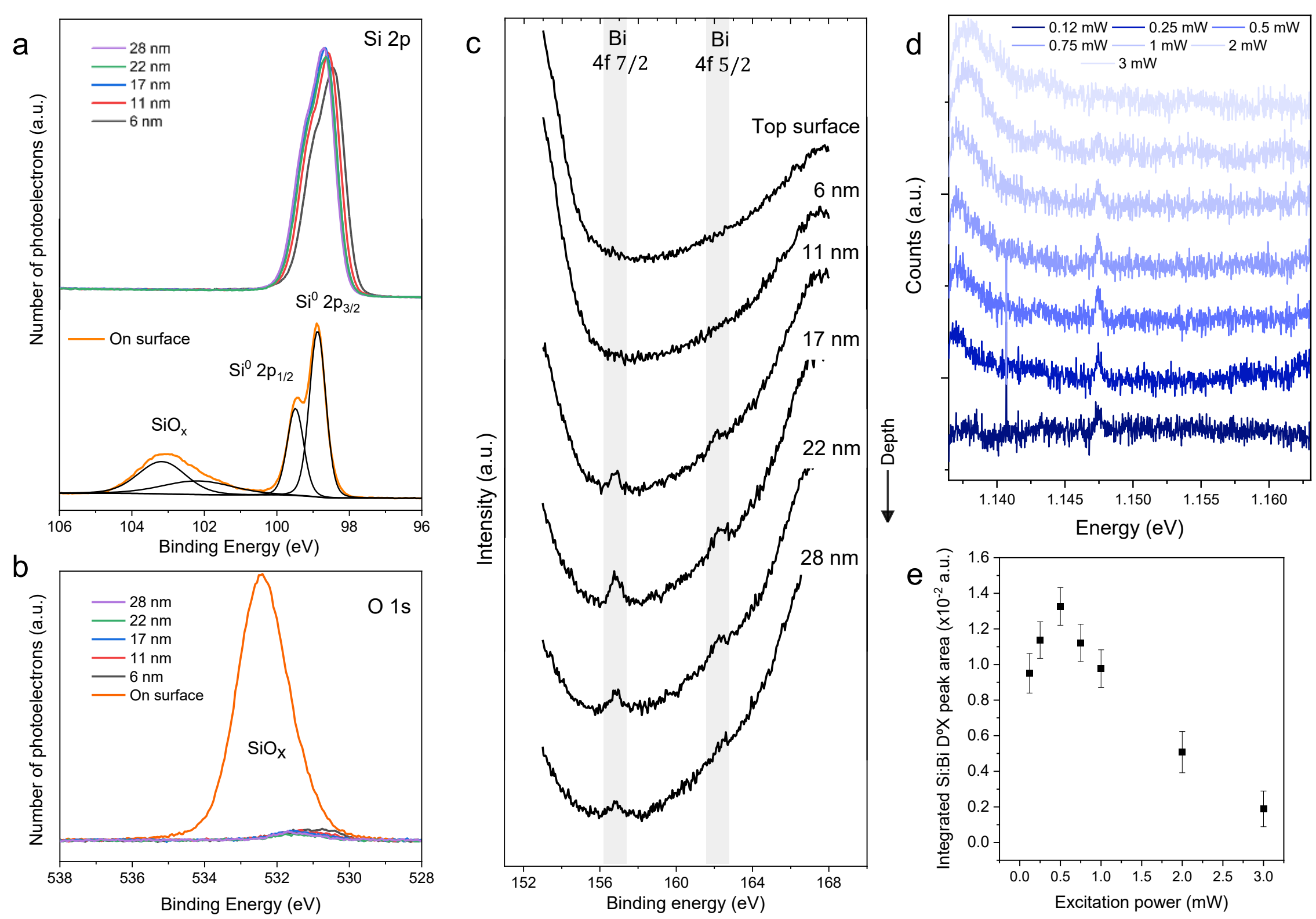

**FIG. 3. Chemical and optical characterization of the Bi-only sample after annealing at 650 °C for 300 s.** Depth-resolved XPS spectra of the (a) Si 2p, (b) O 1s, and (c) Bi 4f regions. The surface Si $2p$spectrum is decomposed into elemental Si and oxidized Si contributions, while the rapid suppression of the O $1s$signal after sputtering indicates that oxidation is primarily confined to the surface. The Bi $4f_{7/2}$ and $4f_{5/2}$ regions are highlighted near 156.7 and 162.4 eV, respectively. Spectra in (c) are vertically offset for clarity. (d) Excitation-power-dependent photoluminescence spectra measured at 6 K, showing the Si:Bi donor-bound-exciton transition near 1.1475 eV. The spectra are vertically offset for clarity. (e) Integrated Si:Bi $D^0X$peak area as a function of excitation power. The emission reaches a maximum near 0.5 mW and decreases at higher power, indicating excitation-induced saturation and quenching. Error bars represent the standard errors obtained from the peak fitting.

The low-temperature (6 K) power-dependent photoluminescence spectra in Fig. 3(d) provide a complementary probe of the donor environment. A weak feature near 1.1475 eV is consistent with the Si:Bi $D^0X$ transition reported for implanted Bi donors [33], providing a donor-related optical signature after annealing at 650 °C. The integrated peak area in Fig. 3(e) increases up to approximately 0.5 mW and then decreases at higher excitation power. This initial increase of the emission followed by a quenching at high-power is consistent with a saturation of the Bi emission due to the finite Bi concentration and competing carrier trapping or nonradiative recombination enabled at high density carrier excitation. Implantation-related defects have similarly been reported to suppress Si:Bi donor-bound-exciton emission in implanted silicon despite the presence of electrically active donors [33]. Additional power-dependent photoluminescence measurements for both implantation routes after extended annealing at 850 and 950 °C are provided in ***Supporting Information Section S4***. The optical measurements establish donor-related activity but do not determine the lattice position or depth distribution of the implanted Bi. We therefore used random and aligned RBS measurements to quantify the substitutional population and examine features associated with Bi redistribution outside the main implantation profile.

Figure 4(a) shows the substitutional Bi areal density extracted from the random and aligned RBS spectra. After annealing at 650 °C, the Bi-only and Si-preimplanted/Bi samples contain comparable substitutional populations, corresponding to fractions of $91 \pm 2\%$ and $89 \pm 4\%$, respectively, within the main implanted region. This similarity contrasts with their markedly different Hall-derived electrical activation yields of approximately 60% and 90%. The substitutional fractions are evaluated within the main implanted Bi region, whereas the Hall-derived activation yield is normalized to the nominal implanted fluence. These quantities therefore use different reference populations and should not be interpreted as identical dose fractions. The higher Hall-derived activation yield observed for the Si-preimplanted/Bi sample at 650 °C therefore cannot be attributed simply to a larger number of Bi atoms occupying substitutional sites. Channeling measurements identify Bi aligned with the Si lattice but do not establish whether each substitutional atom contributes a mobile electron, particularly in the presence of compensating defects, strain, or electrically inactive Bi-related configurations [31,42].

The substitutional Bi density generally decreases with increasing annealing temperature. For the Bi-only route, it decreases from approximately $7.1 \times 10^{13}$ $\text{cm}^{-2}$ at 650 °C to $2.0 \times 10^{13}$ $\text{cm}^{-2}$ at 950 °C. The Si-preimplanted/Bi route exhibits an earlier reduction at 750 °C, followed by a value comparable to that of the Bi-only route near 850 °C. The Hall-derived activation yields subsequently converge at higher temperature, although no 950 °C substitutional-density value is shown here for the Si-preimplanted/Bi route. Thermal-budget-dependent losses of substitutional and electrically active Bi have previously been associated with diffusion, displacement from ideal lattice rows, and the formation of inactive Bi–defect configurations [31,36]. The nonmonotonic evolution of the preimplanted series further indicates that annealing does not produce a simple, uniform removal of Bi from substitutional sites.

Depth-resolved analysis of the Si-preimplanted/Bi sample reveals an additional weak feature outside the main near-surface Bi peak [Fig. 4(b)]. Its energy-loss position corresponds to a depth of approximately 220 nm, at or beyond the former amorphous/crystalline interface and within the end-of-range region produced by the deeper Si preimplantation. The feature is therefore consistent with a buried Bi-related population trapped near residual end-of-range defects, although a contribution from surface As, while unlikely, cannot be completely excluded. Extended defects generated near the end of shallow implantation-damage profiles are known to act as sinks for mobile impurities during annealing [35]. The spatial association of this buried feature with the deeper regrowth pathway supports the conclusion that amorphization depth influences Bi redistribution during solid-phase epitaxial regrowth. This observation further suggests that increasing the preamorphized-layer thickness could provide a route to spatially separate buried Bi-related defects from the near-surface donor region, which may be beneficial for future spin-based device architectures.

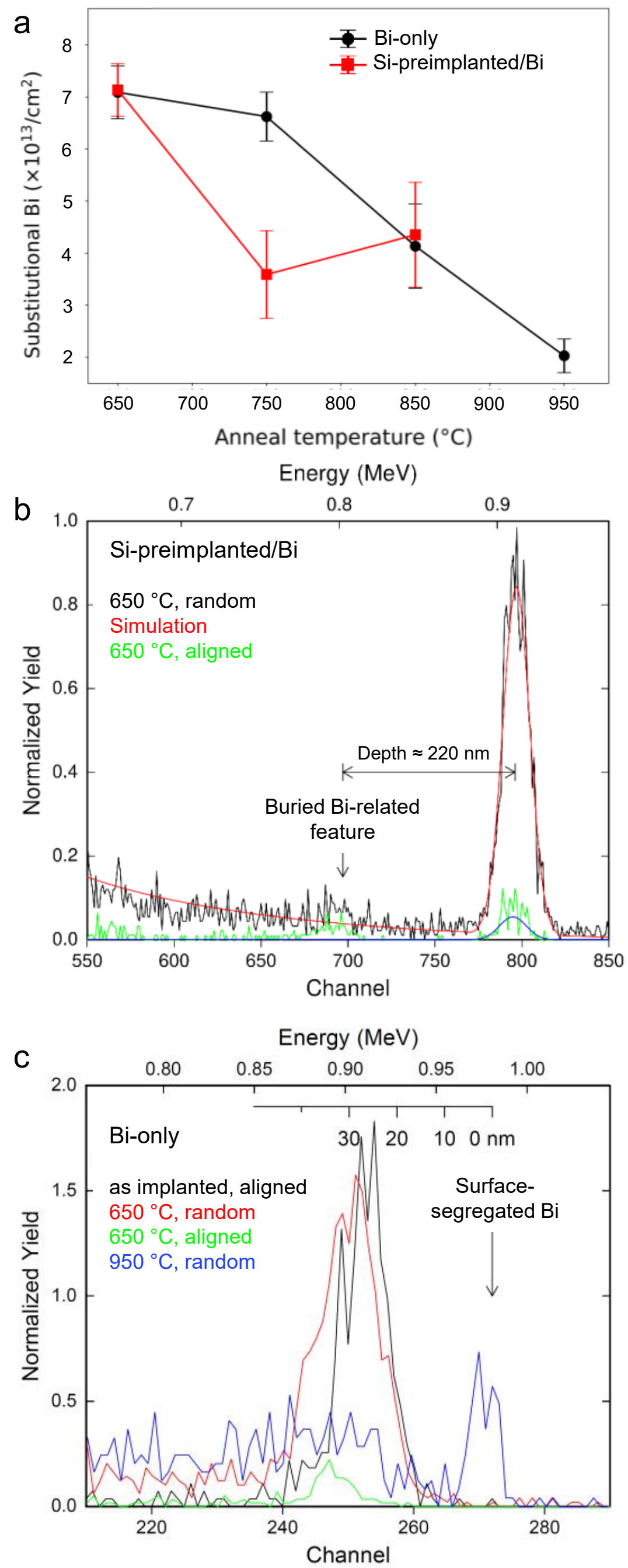

**FIG. 4. Amorphization-depth-dependent Bi incorporation and redistribution.** (a) Substitutional Bi areal density as a function of annealing temperature for the Bi-only and Si-preimplanted/Bi routes. Note that no reliable 950 °C substitutional-density value was obtained for the Si-preimplanted/Bi route (***Supporting Information Section S2***). Lines are guides to the eye. (b) Random and aligned RBS spectra of the Si-preimplanted/Bi sample after annealing at 650 °C, together with the simulated spectrum. The weak feature corresponding to a depth of approximately 220 nm is consistent with a buried Bi-related population at or beyond the former amorphous/crystalline interface, within the end-of-range region. (c) Comparison of the 650 and 950 °C random spectra shows depletion of the main implanted-region signal and the emergence of a distinct contribution at the Bi surface position, consistent with surface segregation.

At higher thermal budgets, redistribution toward the free surface becomes increasingly important. In the Bi-only sample, comparison of the random spectra measured after annealing at 650 °C and 950 °C shows the development of a distinct contribution at the Bi surface position, accompanied by depletion and broadening of the signal from the original implanted region [Fig. 4(c)]. This evolution provides evidence that the reduction in substitutional and electrically active Bi is accompanied by spatial redistribution and surface segregation, rather than solely by a local change in lattice-site occupancy. Diffusion and profile redistribution have likewise been reported for implanted Bi under increased annealing budgets [31,36].

## 3. Conclusion.

In summary, we show that the electrical activation of near-surface Bi donors is influenced by the implantation-induced amorphous-layer geometry and the resulting regrowth pathway, rather than by substitutional incorporation alone. Direct Bi implantation and Si preimplantation produce substantially different amorphous-layer depths of approximately 47 and $177 \pm 15$ nm, respectively. After annealing at 650 °C, both routes recover crystalline order and exhibit comparable substitutional Bi fractions of $91 \pm 2\%$ and $89 \pm 4\%$ within the main implanted region, yet their Hall-derived electrical activation yields differ markedly, reaching approximately 60% and 90%, respectively. Depth-resolved ion scattering further reveals a buried Bi-related feature at or beyond the former amorphous/crystalline interface, within the end-of-range region, consistent with interaction with residual defects. At higher thermal budgets, depletion of the main implanted region and surface segregation accompany the reduction of substitutional and electrically active Bi. XPS and low-temperature photoluminescence measurements on the Bi-only sample reveal no resolvable Bi-oxide contribution and provide a donor-related optical signature. These results demonstrate that the spatial redistribution of Bi during regrowth, together with the initial amorphous-layer geometry, plays a key role in determining the electrically active Bi population in shallow Si:Bi structures for quantum-device applications.

## 4. Experimental Section.

***Sample Preparation.***

High-resistivity, (100)-oriented float-zone Si wafers (4–8 kΩ cm) were used. Two implantation sequences were investigated. In the first, Si wafers were implanted with Bi at 20 keV and a fluence of $1 \times 10^{14}$ $cm^{-2}$ (Bi-only). In the second, the wafers were preamorphized with 60 keV $Si^{+}$ at a fluence of $1 \times 10^{16}$ $cm^{-2}$, followed by the same Bi implantation (Si-preimplanted/Bi). These conditions were selected to generate substantially different implantation-induced amorphization depths while maintaining the same near-surface Bi implantation conditions. All implantations were performed at room temperature with a 7° tilt to suppress ion channeling. The samples were subsequently annealed in $N_2$ for 300 s at 650, 750, 850, or 950 °C. Unimplanted Si wafers from the same starting material were subjected to identical annealing treatments and used as references for the Hall measurements. Bi implantation and damage profiles were simulated using the Stopping and Range of Ions in Matter (SRIM/TRIM) code [37].

***Hall Measurements.***

Hall measurements were performed using a four-probe Van der Pauw configuration. Ti (4 nm)/Au (100 nm) Ohmic contacts were patterned on the sample surface by photolithography, with a spacing of 5 mm between adjacent contacts, and subsequently annealed at 300 °C for 60 s. Electrical measurements were performed at room temperature in the dark while the magnetic field was swept between −3600 and +3600 G. The Hall sheet carrier density was obtained directly from the field-dependent Hall response. For each annealing condition, the sheet carrier density measured for an identically annealed unimplanted Si reference was subtracted from that of the implanted sample. The electrical activation yield was then calculated by normalizing the corrected Hall sheet carrier density to the nominal Bi fluence of $1 \times 10^{14}$ $cm^{-2}$. Electrical resistivity and Hall mobility were evaluated independently as complementary transport parameters. Details of the Van der Pauw analysis and activation-yield calculation are provided in Supporting Information Section S1.

***Secondary Ion Mass Spectrometry.***

Dynamic secondary ion mass spectrometry (SIMS) measurements were performed using a CAMECA SC-Ultra magnetic-sector instrument equipped with a floating-voltage primary column and secondary-ion extraction optics. This configuration enables low primary-ion impact energies and high depth resolution. SIMS was used to determine the as-implanted Bi depth distribution, with a depth resolution on the order of 5 Å.

***X-ray Photoelectron Spectroscopy.***

X-ray photoelectron spectroscopy (XPS) measurements were performed using a Kratos Axis Ultra DLD system with a monochromated Al Kα source (hν = 1486.7 eV). The binding-energy scale was calibrated using the Au $4f_{7/2}$ peak of metallic Au at 83.96 eV. Samples were electrically grounded, and a charge neutralizer was used to minimize charging effects. The spectra were analyzed using CasaXPS (version 2.3.18) and quantified using the relative sensitivity factors provided by Kratos Analytical. Depth profiling was performed by sequential Ar-ion sputtering, and the sputtering rate was calibrated by atomic force microscopy.

***Photoluminescence Measurements.***

Photoluminescence (PL) measurements were performed using a solid-state laser emitting at 532 nm. Samples were mounted in a closed-cycle He cryostat and PL was measured at 6 K. PL spectra were analyzed with a 550-mm spectrometer equipped with 600 gr/mm grating and a 1024-pixel InGaAs camera. The spectral resolution is 0.2 meV.

***Rutherford Backscattering Spectrometry and Channeling.***

Rutherford backscattering spectrometry (RBS) measurements were performed using 1 MeV He ions in both random and Si(100)-aligned geometries. Two detectors were used: a near-backscattering detector positioned 15° from the exact backscattering direction and a glancing-angle detector for which the outgoing ion trajectory was 82.5° from the sample-surface normal. For channeling measurements, the sample orientation was adjusted to minimize the backscattering yield along the Si(100) axis. Random spectra were acquired with the sample tilted by approximately 7° away from the channeling direction while continuously rotating the sample about its surface normal. The detector-specific entrance and exit geometries were taken into account when converting backscattered-ion energy loss into depth. Random and aligned spectra were used to evaluate implantation-induced lattice disorder, Bi redistribution, and the substitutional Bi population.

***Raman Spectroscopy.***

Raman measurements were performed using an AIST-NT OmegaScope system equipped with a Nanofinder 30 Raman spectrometer and a thermoelectrically cooled Andor CCD detector. A 532.1 nm solid-state laser (Cobolt 04-01 series) was focused onto the sample using a 100× Mitutoyo M Plan Apo objective with a numerical aperture of 0.7. The Raman spectra were used qualitatively to compare implantation-induced amorphization and lattice recovery between the Bi-only and Si-preimplanted/Bi samples.

**SUPPORTING INFORMATION**

# Amorphization-Depth-Dependent Bismuth Redistribution and Activation in Silicon for Quantum Devices

Youcef A. Bioud [1], Simon-Jean Michel [2], Sebastien Francoeur [2], Sjoerd Roorda [3], Eva Dupont-Ferrier [1,4]

[1] *Institut Quantique, Université de Sherbrooke, Sherbrooke, Quebec J1K 2R1, Canada*
[2] *Department of Engineering Physics, Polytechnique Montreal, Montreal, QC H3T 1J4, Canada*
[3] *Département de physique, Université de Montréal, Montréal, Québec H3C 3J7, Canada*
[4] *Department of Physics, Université de Sherbrooke, Sherbrooke, Québec J1K 2R1, Canada*

Corresponding authors: y.bioud@usherbrooke.ca , eva.dupont-ferrier@usherbrooke.ca

**S1. Hall and Van der Pauw Electrical Measurements**

Electrical transport measurements were performed in a four-probe Van der Pauw geometry. Identically annealed unimplanted Si wafers were measured under the same conditions and used as references for the background carrier contribution. The Bi electrical activation yield was determined from the Hall sheet carrier density, expressed directly in $cm^{-2}$; therefore, no electrically active layer thickness is required for the activation-yield calculation.

**Van der Pauw and Hall analysis.**

For two characteristic resistances, $R_A$ and $R_B$, measured in the four-probe geometry, the sheet resistance $R_s$ satisfies the Van der Pauw relation:

$$exp(-\frac{\pi R_A}{R_s}) + exp(-\frac{\pi R_B}{R_s}) = 1$$

The sheet Hall coefficient was obtained from the field-dependent Hall voltage:

$$R_{H,s} = \frac{\Delta V_H}{I\ \Delta B}$$

Using the magnitude of the Hall coefficient, the corresponding Hall sheet carrier density is

$$N_s = \frac{1}{q\ |R_{H,s}|}$$

where *q* is the elementary charge. Using the same magnitude convention, the Hall mobility may be written in sheet form as

$$\mu_H = \frac{|R_{H,s}|}{R_s}$$

The electrical resistivity and Hall mobility are reported as complementary transport parameters and are not used to calculate the Bi activation yield.

**Background correction and electrical activation yield.**

For each annealing temperature *T,* the Hall sheet carrier density measured for the identically annealed unimplanted Si reference was subtracted from that of the corresponding implanted sample:

$$N_{s,corr}(T) = N_{s,implanted}(T) - N_{s,reference}(T)$$

The corrected Hall sheet carrier density was then normalized to the nominal implanted Bi fluence

$$\Phi_{Bi} = 1.0 \times 10^{14}\ cm^{-2}$$

according to

$$Y_{act}(T) = \frac{N_{s,corr}(T)}{\Phi_{Bi}} \times 100\%$$

or, explicitly,

$$Y_{act}(T) = \frac{N_{s,implanted}(T) - N_{s,reference}(T)}{1.0 \times 10^{14}\ cm^{-2}} \times 100\%$$

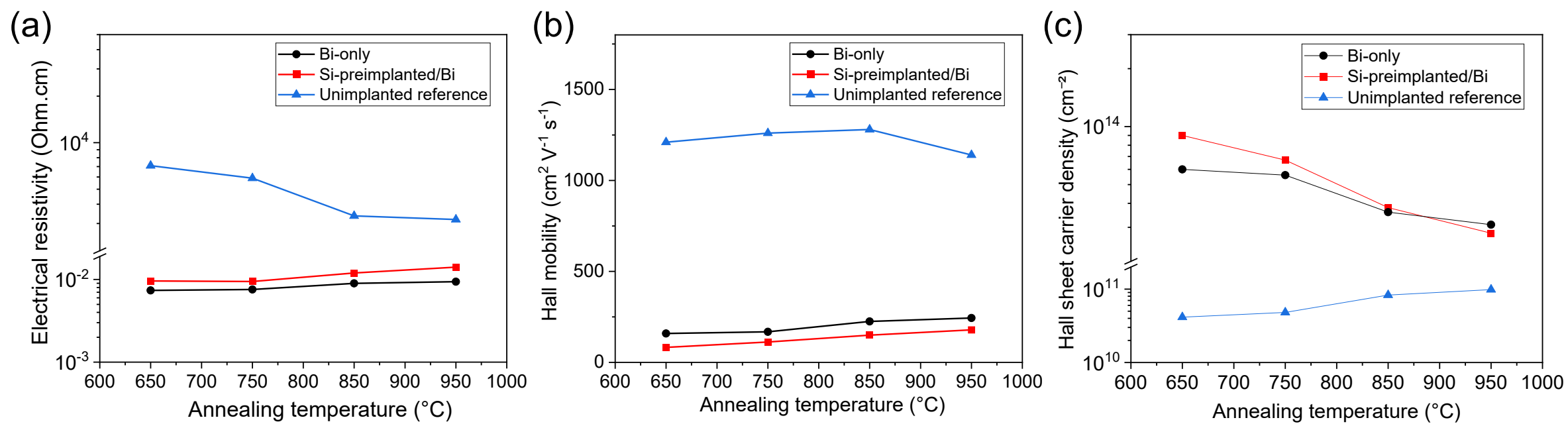

**FIG. S1.** Electrical transport properties of Bi-implanted Si. (a) Electrical resistivity, (b) Hall mobility, and (c) Hall sheet carrier density as a function of annealing temperature for the Bi-only, Si-preimplanted/Bi, and identically annealed unimplanted Si reference samples. Lines connecting the experimental data points are guides to the eye.

**TABLE S1.** Hall sheet carrier densities and Bi electrical activation yields. The implanted-sample values are background-corrected using the identically annealed unimplanted Si reference.

| Anneal (°C) | Unimplanted reference $N_s$ ($cm^{-2}$) | Bi-only $N_s$ ($cm^{-2}$) | Bi-only Yact (%) | Si-preimplanted/Bi $N_s$ ($cm^{-2}$) | Si-preimplanted/Bi Yact (%) |
|---|---|---|---|---|---|
| 650 | $4.15 \times 10^{10}$ | $6.0 \times 10^{13}$ | 60 | $9.0 \times 10^{13}$ | 90 |
| 750 | $4.82 \times 10^{10}$ | $5.6 \times 10^{13}$ | 56 | $6.7 \times 10^{13}$ | 67 |
| 850 | $8.30 \times 10^{10}$ | $3.6 \times 10^{13}$ | 36 | $3.8 \times 10^{13}$ | 38 |
| 950 | $9.86 \times 10^{10}$ | $3.1 \times 10^{13}$ | 31 | $2.8 \times 10^{13}$ | 28 |

**Example calculation at 650 °C.**

For the Bi-only sample, the background-corrected Hall sheet carrier density is

$$N_s = 6.0 \times 10^{13}\ cm^{-2}$$

and therefore

$$Y_{act} = \frac{6.0 \times 10^{13}\ cm^{-2}}{1.0 \times 10^{14}\ cm^{-2}} \times 100\% = 60\%$$

For the Si-preimplanted/Bi sample,

$$N_s = 9.0 \times 10^{13}\ cm^{-2}$$

which gives

$$Y_{act} = \frac{9.0 \times 10^{13}\ cm^{-2}}{1.0 \times 10^{14}\ cm^{-2}} \times 100\% = 90\%$$

The electrical activation yield was calculated from the background-corrected Hall sheet carrier density and therefore does not require an assumed electrically active layer thickness. Although the unimplanted Si contribution is subtracted, residual implantation damage may still influence the Hall response through compensation, trapping, mobility variations, or parallel conduction. The reported values are therefore interpreted as Hall-derived electrical activation yields.

The electrical resistivity in Fig. S1(a) was obtained from the measured sheet resistance using the physical wafer thickness,

$$\rho = R_s\, d$$

where $R_s$ is the sheet resistance and *d* is the physical thickness of the Si wafer. The resulting value is reported only as a complementary effective sample resistivity and is not used in the calculation of the Bi electrical activation yield.

## S2. Additional RBS/Channeling Analysis

Rutherford backscattering spectrometry (RBS) measurements were performed using 1 MeV He ions in random and Si(100)-aligned geometries. Spectra were acquired using a near-backscattering detector and a glancing-angle detector, the latter providing enhanced depth sensitivity. Random spectra were used to determine the total Bi areal density, whereas comparison of random and aligned Bi yields was used to estimate the substitutional Bi population. The conversion from energy loss to depth accounts for the different entrance and exit geometries of the two detectors. Because annealing-induced redistribution and detector geometry affect the quantitative analysis, particularly at the highest annealing temperatures, the RBS-derived values are interpreted together with the corresponding spectral profiles shown in the main text. For the Si-preimplanted/Bi sample annealed at 950 °C, the strong Bi redistribution prevents a sufficiently reliable quantitative extraction of the substitutional Bi areal density; therefore, no value is reported.

## S3. XPS Chemical-State and Depth-Profile Analysis

**TABLE S2.** XPS-derived chemical composition and core-level parameters of the Bi-only sample annealed at 650 °C for 300 s as a function of sputter depth.

| Etch Time (s) | Estimated depth (nm) | Core level | Binding energy (eV) | FWHM (eV) | Atomic concentration (at.%) |
|---|---|---|---|---|---|
| **0** | 0 | O 1s | 532.4 | 1.7 | 44.3 |
| | | C 1s | 284.6 | 1.6 | 4.2 |
| | | Si 2p | 98.9 | 0.9 | 51.5 |
| **30** | 6 | O 1s | 530.8 | 1.9 | 3.9 |
| | | Si 2p | 98.5 | 1.1 | 96.1 |
| **60** | 11 | O 1s | 531.5 | 1.6 | 1.7 |
| | | Bi 4f 7/2 | 156.7 | 0.7 | 0.03 |
| | | Si 2p | 98.6 | 1.1 | 98.2 |
| **90** | 17 | O 1s | 531.4 | 0.9 | 1.3 |
| | | Bi 4f 7/2 | 156.6 | 0.5 | 0.03 |
| | | Si 2p | 98.7 | 1.1 | 98.7 |
| **120** | 22 | O 1s | 531.9 | 0.3 | 0.9 |
| | | Bi 4f 7/2 | 156.5 | 0.5 | 0.02 |
| | | Si 2p | 98.7 | 1.1 | 99.0 |
| **150** | 28 | O 1s | 531.1 | 1 | 1.0 |
| | | Bi 4f 7/2 | 156.8 | 0.2 | 0.02 |
| | | Si 2p | 98.7 | 1.1 | 99.0 |

**Note.** The sputter depths were estimated from the calibrated sputtering rate. The Bi 4f7/2 signal becomes detectable within the implanted region between approximately 11 and 28 nm. Because of the weak Bi signal and the overlap of reported binding energies for different non-oxidized Bi configurations, the XPS measurements alone do not determine the lattice-site occupancy of Bi.

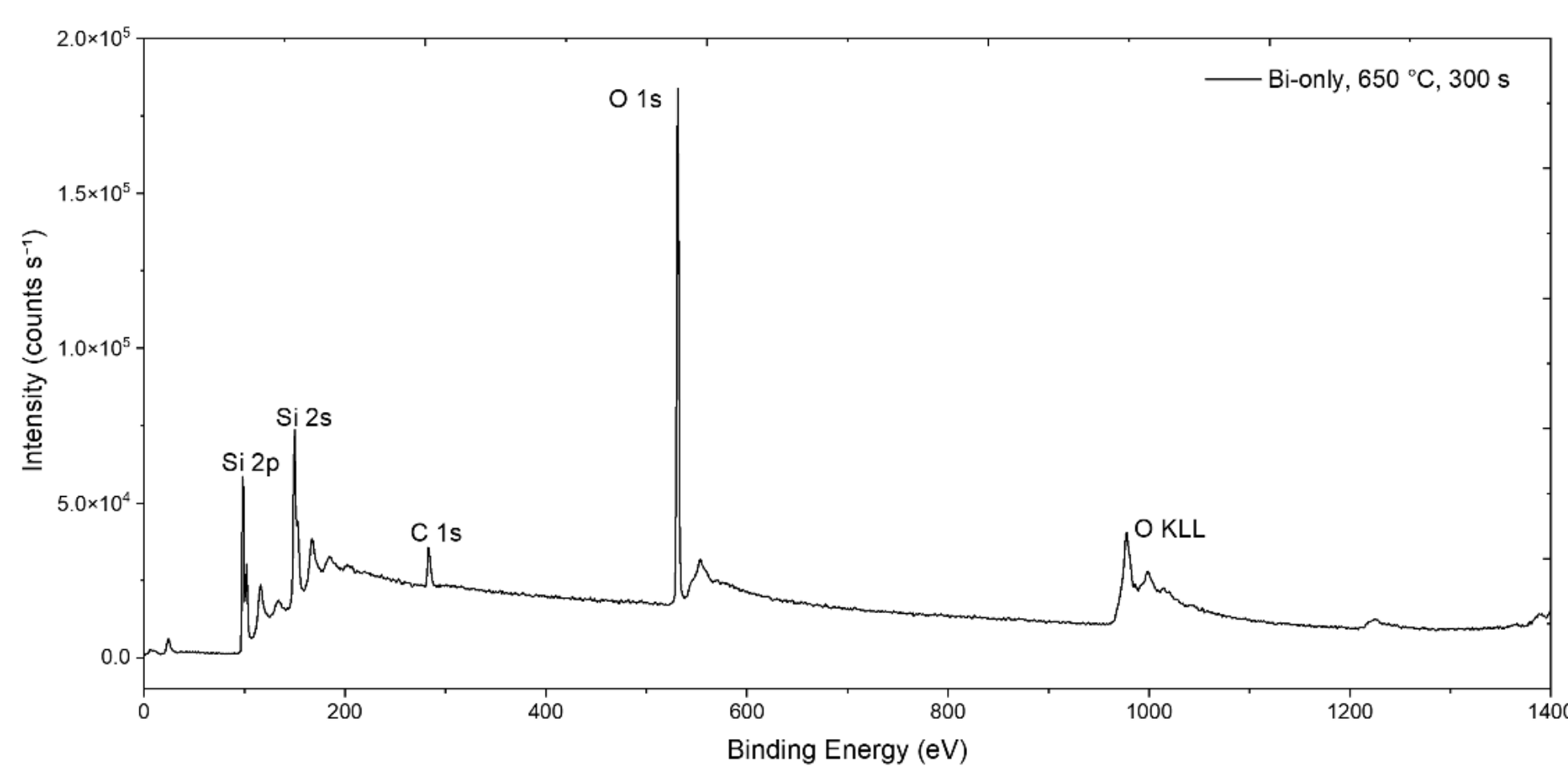

**FIG. S2.** XPS survey spectrum of the Bi-only sample annealed at 650 °C for 300 s. The principal Si, C, and O photoelectron and Auger features are indicated. The weak Bi signal is analyzed using high-resolution Bi 4f spectra in Fig. 3 and Table S2.

## S4. Additional Photoluminescence Measurements

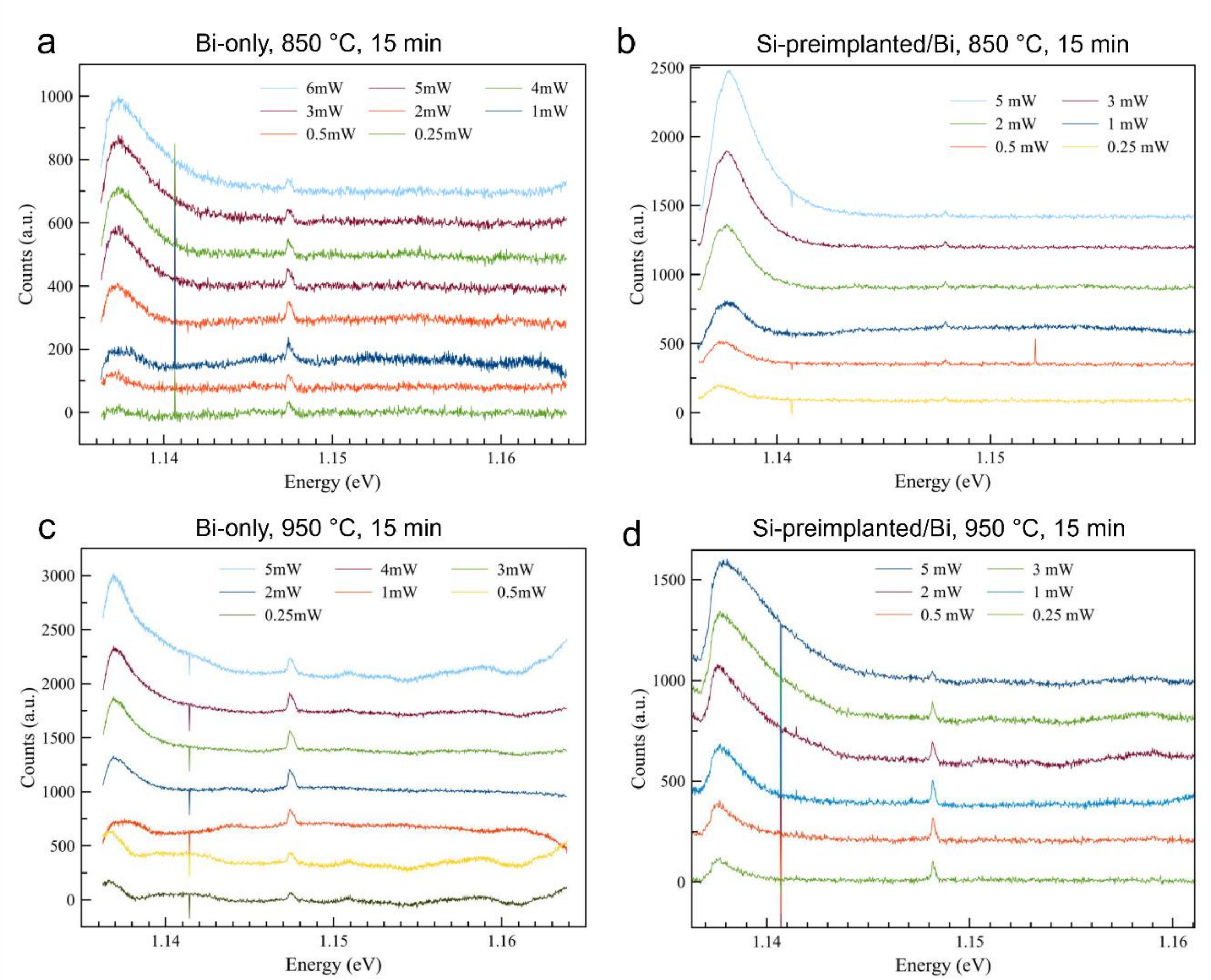

**FIG. S3.** Excitation-power-dependent photoluminescence spectra of Bi-implanted Si after extended thermal annealing. (a,b) Bi-only and Si-preimplanted/Bi samples annealed at 850 °C for 15 min, respectively. (c,d) Corresponding spectra after annealing at 950 °C for 15 min. Measurements were performed at low temperature under excitation powers ranging from 0.25 to 5–6 mW. The expected Si:Bi donor-bound-exciton transition near 1.1475 eV is indicated. Spectra are vertically offset for clarity.

**Note.** These measurements were performed after an extended annealing time of 15 min and are therefore presented as complementary optical characterization rather than as a direct continuation of the 300 s annealing series discussed in the main text.